\documentclass[prl,superscriptaddress]{revtex4-2}

\usepackage{amsmath} 
\usepackage{amssymb}  
\usepackage{amsfonts} 
\usepackage{dsfont}
\usepackage{bm}  
\usepackage{bbm}   
\usepackage{bbold}   
\usepackage{braket} 
\usepackage{color} 
\usepackage{comment}  
\usepackage{dcolumn}  
\usepackage{enumerate}  
\usepackage{epsfig}  
\usepackage{gensymb}  
\usepackage{graphicx}  
\usepackage{indentfirst}  \usepackage{lmodern}  
\usepackage{mathrsfs}  
\usepackage{mathtools}  
\usepackage{psfrag}  
\usepackage{pst-all} 
\usepackage{soul}  
\usepackage{xcolor}
\usepackage{float} 
\usepackage[colorlinks,linkcolor=blue,citecolor=blue,urlcolor=blue,hyperindex,driverfallback=dvipdfm]{hyperref}  \usepackage[T1]{fontenc} 

\DeclareMathOperator{\sgn}{sign}

\usepackage{dirtytalk}
\usepackage{CJKutf8}

\usepackage{float} 
\usepackage{placeins}
\usepackage{bm}
\usepackage{enumitem}

 \usepackage[normalem]{ulem} 
 \makeatletter
\newcommand*{\addFileDependency}[1]{
  \typeout{(#1)}
  \@addtofilelist{#1}
  \IfFileExists{#1}{}{\typeout{No file #1.}}
}
\makeatother

\usepackage[strict]{changepage}
\usepackage{calc}
\usepackage{dirtytalk}
\usepackage{pifont}

\usepackage{xr-hyper} 
\usepackage{wrapfig}

\usepackage{graphicx}
\usepackage{cancel}
\usepackage{longtable}
\usepackage{array}
\newcolumntype{C}[1]{>{\centering\arraybackslash}p{#1}}

\usepackage{tensor}

\usepackage{ulem}
\usepackage{titlesec}
 \titleformat{\paragraph}[hang]{\bfseries}{}{0pt}{\uline}
\titlespacing*{\paragraph}
{0pt}{3.25ex plus 1ex minus .2ex}{1.5ex plus .2ex}

\usepackage{dirtytalk}
\usepackage{cleveref}
\usepackage[]{siunitx}
\usepackage{textcomp}
\usepackage{upgreek}
\begin{document}

\title{Toroidal pulse enhanced XUV generation}
\author{Lu Wang*}
\affiliation{Department of Physics, University of Ottawa, Ottawa, Ontario K1N 6N5, Canada}
\email{lu.wangTHz@outlook.com}
\author{Clément Lacoste}
\affiliation{Department of Physics, University of Ottawa, Ottawa, Ontario K1N 6N5, Canada}
\author{Paul Corkum}
\affiliation{Department of Physics, University of Ottawa, Ottawa, Ontario K1N 6N5, Canada}
\author{Thomas Brabec}
\affiliation{Department of Physics, University of Ottawa, Ottawa, Ontario K1N 6N5, Canada}

\author{Zenghu Chang}
\affiliation{Department of Physics, University of Ottawa, Ottawa, Ontario K1N 6N5, Canada}

\begin{abstract}
Extreme ultraviolet (XUV) radiation plays a central role in a wide range of applications. Despite being a cornerstone of attosecond science, efficient XUV generation remains highly challenging. Here, we investigate XUV generation via nonlinear Thomson scattering driven by a toroidal pulse. Unlike conventional laser pulses, toroidal pulses possess intrinsically non-separable spatiotemporal field distributions. This unique field structure produces strongly asymmetric electron acceleration, resulting in substantially enhanced XUV emission that is 2–4 orders of magnitude stronger than that generated by a Gaussian pulse. Our results identify toroidal pulses as a promising route toward compact ultrafast XUV sources and advanced strong-field light–matter interaction studies.
\end{abstract}

\maketitle

\section{Introduction}
Extreme ultraviolet (XUV) radiation finds numerous applications, such as time-resolved photoelectron spectroscopy \cite{boschini2024time}, lithography \cite{wu2007extreme}, ultrafast spin control \cite{londo2022ultrafast}, and 
probing ultrafast atomic/molecular dynamics/chirality \cite{sansone2010electron,itatani2004tomographic,smirnova2009high,beaulieu2017attosecond}. Although being a cornerstone of attosecond science, obtaining efficient coherent XUV radiation is extremely challenging. XUV pulses are typically generated via high-harmonic generation, a highly nonlinear optical process that requires an intense laser to trigger emission at integer multiples of the driving-field frequency. Because available high-energy ultrafast lasers are primarily in the infrared, generating XUV radiation through high-harmonic generation typically requires upconversion of several tens of harmonic orders, which is inherently inefficient. Although high-flux XUV sources are available from synchrotrons and free-electron lasers, these facilities are generally large-scale, costly, energy-intensive, and accessible only to a limited number of users.

Apart from standard high-harmonic generation, where emission arises from the recombination of an ionized electron with its parent ion, a free electron interacting with an intense laser and reaching relativistic speed can also emit radiation in the XUV or beyond. This radiation, known as relativistic nonlinear Thomson scattering \cite{lee2003relativistic}, arises from the interaction of a free electron with the driving laser’s electric and magnetic fields, both of which play essential roles in shaping the emitted radiation.

Regardless of the high-harmonic generation or the nonlinear Thomson scattering, most conventional laser-driven XUV generation schemes primarily rely on increasing the laser intensity. Extensive studies have explored radiation driven by linearly/circularly polarized Gaussian pulses, radially/azimuthally polarized beams, and optical vortex beams \cite{kong2019vectorizing,lee2003relativistic,gariepy2014creating}. Despite their diverse optical characteristics, these pulses generally possess separable spatial and temporal profiles. In other words, the electric field can be expressed by a product of a spatial function and a temporal function. In contrast, electromagnetic pulses with intrinsically non-separable spatiotemporal structures also exist. In 1996, Hellwarth and Nouchi theoretically introduced a fundamentally different class of electromagnetic excitation known as the toroidal pulse \cite{hellwarth1996focused}. More recently, experimental endeavors in generating toroidal pulses at both terahertz and optical frequencies have renewed significant interest in their unique electromagnetic properties and light–matter interactions \cite{zdagkas2022observation,wang2024observation,jana2024quantum}.

Toroidal pulses exhibit a distinct toroidal topology: the magnetic field forms a doughnut-like structure around the propagation axis, while the electric field wraps along the doughnut surface, leading to a pronounced longitudinal component aligned with the direction of propagation. Thus, the toroidal pulse is also called the "flying electromagnetic doughnut" \cite{zdagkas2022observation}. At the focal center, the toroidal pulse is characterized by a tightly localized, single-cycle electric field \cite{hellwarth1996focused}. As a topologically structured light field with coupled ultrafast temporal and vectorial spatial features, the toroidal pulse can support unusual electromagnetic excitations in matter, including singularities, vortex rings, and optical skyrmionic structures \cite{wang2025hybrid,shen2021supertoroidal}.

In this work, we investigate XUV generation via nonlinear Thomson scattering driven by a toroidal pulse. Our results show that, unlike conventional laser pulses, the distinctive spatiotemporal coupling of toroidal fields enables more efficient electron acceleration and significantly enhances the emitted radiation. In addition, we developed a fully analytical theoretical framework to characterize toroidal pulses, allowing their pulse parameters to be systematically compared with those of other well-known optical fields. These findings open a new pathway toward compact ultrafast XUV sources and advanced studies of strong-field light–matter interactions.

\section{Results}
To highlight the distinctive properties of the toroidal pulse, we compare the radiation emitted by a single electron driven by toroidal, Gaussian, and radially polarized pulses, with the latter two serving as standard benchmarks for particle radiation. The mathematical definition of the electric and magnetic fields of these three cases can be found in the Supplementary Material Eqs.(S37), Eqs.(S50-55), and Eqs.(S89-S91), respectively. For a fair comparison, all three pulses are chosen to have the same pulse energy $\sim 22\,\mathrm{mJ}$, beam waist $W_0 \sim 15\,\mu\mathrm{m}$, effective central frequency $\omega_c \sim 0.31\,\Omega_0$ ($\Omega_0 = {2\pi c}/{\lambda_0}$, $\lambda_0 = 0.8\,\mu\mathrm{m}$), and effective pulse duration is $\tau_c = 4.1\,\mathrm{fs}$. This also leads to identical peak electric field strength $\sim 2\times 10^{12}\,\text{V/m}$. Further details are provided in the Methods section and Supplementary Material Section V.

In addition to the laser parameters, electron dynamics in electromagnetic fields are determined by two initial conditions: position and velocity. In this work, we consider an electron initially at rest, $\bm{v}_0=0$, so that all motion and radiation arise solely from the applied field, enabling a clearer interpretation of the underlying physics. The initial position is chosen for each pulse: $y=0$, and $x$ is set to the location that yields the highest radiation. As a general guideline, the optimal position along $x$ lies near the location of the peak field strength: $x=0$ for a Gaussian beam, and near the center of the doughnut-shaped intensity ring for the other two cases (see Supplementary Material Fig. S4). With these choices, the radiation along the $y$ dimension is zero for all three cases.

\begin{figure*}[h]
\centering
\includegraphics[width=1\linewidth]{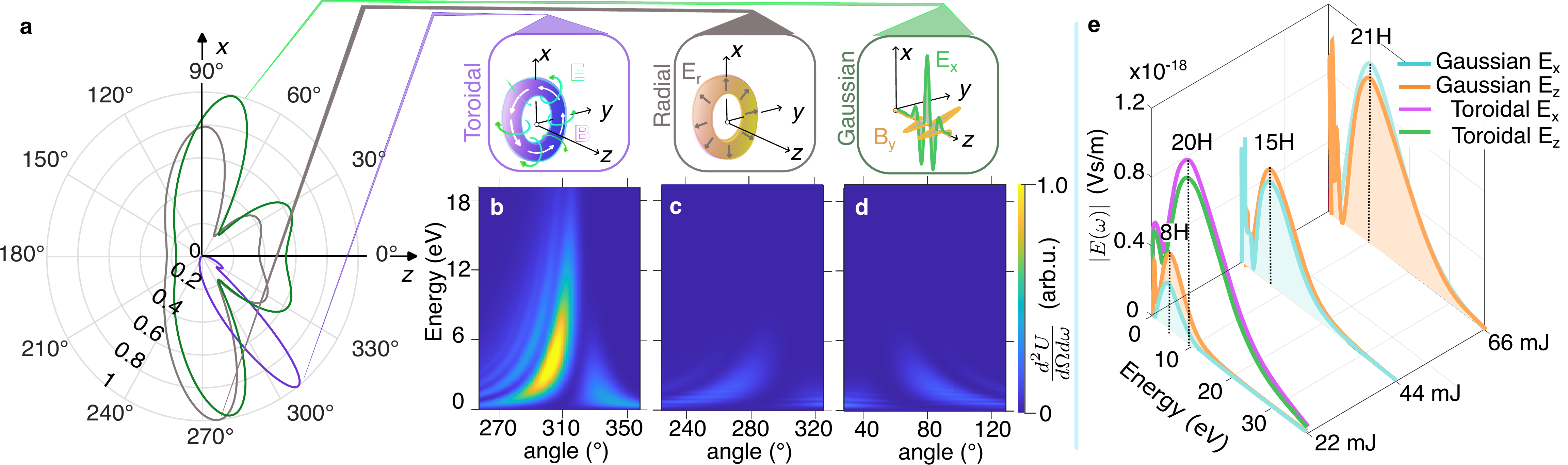}
\caption{Radiation emission from a single electron driven by a toroidal pulse, a radially polarized pulse, and a Gaussian beam, respectively. The laser parameters are identical for all cases: total pulse energy $\sim 22\,$mJ, effective pulse duration $\tau_c=4.1\,$fs (single-cycle pulse), effective wavelength $\lambda_c=\lambda_0/0.31$ with $\lambda_0=0.8$\textmu m, and beam waist $W_0=15\,\mu$m. Panel \textbf{a} shows the angular distribution of the radiated energy from a single in the $x$–$z$ plane, where the $z$-axis denotes the laser propagation direction.  Panels \textbf{b–d} display the corresponding radiation spectral-angular distribution $dU/(d\omega d\Omega)$ evaluated near the angle of maximum emission identified in panel \textbf{a} for each case. Here, $U$ is the emitted energy, $\omega$ is the angular frequency, and $\Omega$ is the solid angle. Panel \textbf{e} compares the spectra of the Gaussian beam and the toroidal pulse at their respective peak-emission angles. Vertical dashed lines indicate the spectral maxima, with the corresponding upconverted harmonic orders of $\omega_c$ denoted alongside. The Gaussian spectra are obtained using pulse energies $22 \,$mJ, $44\,$mJ, and $66\,$mJ, respectively. As the input energy increases, the Gaussian beam requires approximately three times higher energy to achieve a spectrum comparable to that produced by the toroidal pulse.}\label{fig1}
\end{figure*}

Figure~\ref{fig1} shows that the toroidal pulse outperforms both radially polarized and Gaussian pulses in electron acceleration. The very comparable results between the radially polarized and the Gaussian beam are because we have used a relatively large beam waist $W_0=15$\,\textmu m to ensure sufficient overlap between the laser field and the electron trajectory. Since the laser pulses are not tightly focused, the longitudinal electric field strength of the radially polarized pulse is much smaller than the radial electric field strength, i.e., $E_z<E_r$ (See Supplementary Material Fig.S3). 

The narrower angular distribution of the toroidal pulse in Fig.~\ref{fig1}\textbf{a} indicates emission from a strongly relativistic electron. The spectral–angular distribution, $dU/(d\omega d\Omega)$, evaluated near the angle of maximum emission identified in panel \textbf{a}, is shown for each case in Fig.~\ref{fig1}\textbf{b–d}. Here, $U$ is the emitted energy, $\omega$ is the angular frequency, and $\Omega$ is the solid angle.
Notably, the radiation generated by the toroidal pulse (Fig.~\ref{fig1}\textbf{b}) extends up to $\sim20$\,eV in the XUV range. Furthermore, Figs.~\ref{fig1}\textbf{b–d} indicate that the emission driven by the toroidal pulse exhibits an energy spectral bandwidth approximately four times wider than that of the Gaussian and radially polarized pulses. In addition, we examine the spectrum of the emitted electric field at the emission angle corresponding to the maximum frequency extent (Fig.~\ref{fig1}\textbf{e}), where the frequency extent is defined by Eq.~(\ref{eq:w_r}) in the Method section. For single-particle emission, this angle is very close to that corresponding to the maximum emission energy, i.e., the selected tip in Fig.~\ref{fig1}\textbf{a}. It can be seen that, for the same pulse energy ($22$\,mJ), the toroidal pulse produces radiation with a much wider spectrum and significantly higher field strength. To achieve a comparable radiation spectrum, the Gaussian beam requires approximately three times more energy.

This naturally raises the question: if all laser parameters are comparable, what enables the toroidal pulse to substantially outperform the others? Our first suspect is the ponderomotive force. This force is proportional to the field intensity and always points from regions of high intensity to low intensity, regardless of the charge of the particle. In other words, the particle is driven out of high-intensity regions. For a Gaussian beam, the ponderomotive force drives the electron transversely toward the edges of the beam. In contrast, toroidal and radially polarized beams exhibit a doughnut-shaped intensity profile, where the ponderomotive force can induce a focusing effect, driving electrons toward the beam center and thereby increasing the interaction length. However, this is not the primary mechanism behind the enhanced radiation of the toroidal pulse. For the chosen parameters, the radially polarized beam shows little difference from the Gaussian beam, indicating that focusing alone cannot explain the observed enhancement.

We have found two reasons for the enhanced radiation from the toroidal pulse. The first reason lies in the electric field form. While Gaussian and radially polarized beams exhibit a cosine-type waveform with a single maximum, the toroidal pulse features a $\sin{}$-type waveform that connects a maximum to a minimum of equal amplitude but opposite sign. This leads to a more rapid variation in the field, which favors electron acceleration for the following reason. Although for Gaussian pulses with the same energy, the $\sin{}$-type waveform has a lower peak field strength, it still extends the radiation spectrum by $\sim 30\%$ compared to the cosine-type waveform (see Supplementary Material Fig.S5). However, this alone is still insufficient to explain the substantial enhancement in both spectral range and emission amplitude observed for the toroidal pulse. 

\begin{figure*}[h]
\centering
\includegraphics[width=1\linewidth]{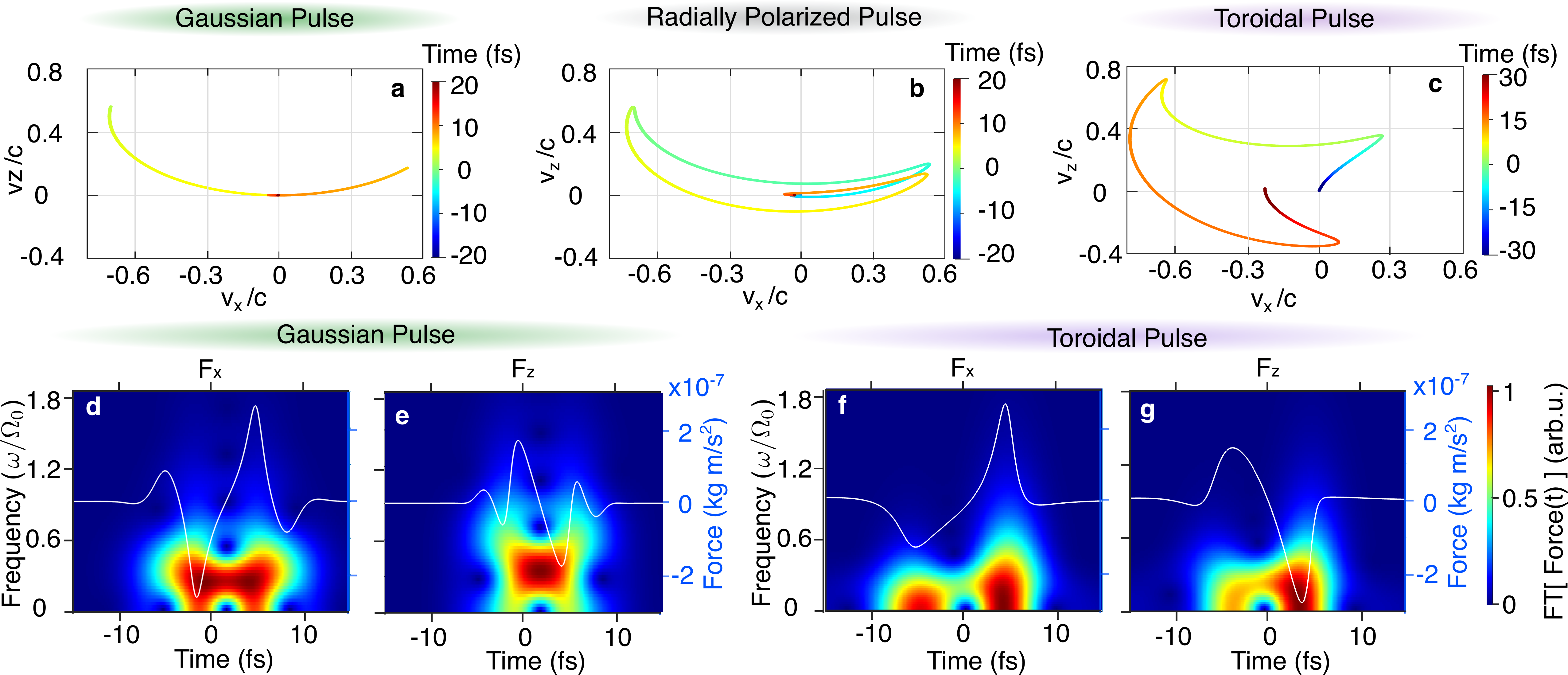}
\caption{The dynamics of velocities along $x$ and $y$ normalized by the speed of light $c$ are presented in panels \textbf{a-c}, where each panel corresponds to Gaussian, radially polarized, and toroidal pulses, respectively. Panels \textbf{d-g} show the windowed Fourier transform of the electromagnetic force experienced by the electron. Panels \textbf{d,e} present the force components along $x$ and $z$, respectively, for the Gaussian pulse, while panels \textbf{f,g} show the corresponding components for the toroidal pulse. The spectral amplitude is shown on the left axis, whereas the time-domain force, indicated by white curves, is plotted on the right axis.
 }\label{fig2}
\end{figure*}

We now turn to the second, and the fundamental reason why the toroidal pulse leads to enhanced nonlinear emission: its non-separable spatiotemporal structure, namely a spatially varying spectrum. To analyze the acceleration dynamics, we present the electron velocity evolution driven by sine-type Gaussian, sine-type radially polarized, and toroidal pulses in Fig.~\ref{fig2}\textbf{a–c}, respectively, with color indicating time. Same as in Fig.~\ref{fig1}, the initial conditions are $\bm{v}_0=0$, $y = 0$, and $x$ is set to the location that yields the highest radiation.

As shown in Fig.~\ref{fig2}\textbf{a–c}, the Gaussian and radially polarized pulses yield negligible net velocity after the interaction, whereas the toroidal pulse produces a substantial net velocity. This indicates that, for the Gaussian and radially polarized pulses, acceleration and deceleration are balanced and cancel over the course of the interaction. In contrast, this cancellation does not occur for the toroidal pulse.

To further investigate this non-balanced interaction, we perform a windowed Fourier transform of the electromagnetic force experienced by the electron driven by the Gaussian and the toroidal pulses (Fig.\ref{fig2}\textbf{d-g}). Results for the radially polarized pulse are similar to the Gaussian case and are therefore omitted here ( see Supplementary Material Fig.~S6). Fig.~\ref{fig2}\textbf{d,e} present the force components along $x$ and $z$, respectively, for the Gaussian pulse, while Fig.~\ref{fig2}\textbf{f,g} show the corresponding components for the toroidal pulse. Since the windowed Fourier transform reflects how the frequency components distribute in time, Fig.~\ref{fig2}\textbf{d,e} reveals a force profile that is highly symmetric about its center, whereas Fig.~\ref{fig2}\textbf{f,g} displays strong asymmetry.

This asymmetry creates an imbalance between acceleration and deceleration, resulting in a large net velocity gain after the interaction. In particular, Fig.~\ref{fig2} suggests that the enhanced nonlinear emission driven by the toroidal pulse originates from two effects. First, the toroidal pulse produces a larger forward velocity $+v_z$, increasing the interaction time between the electron and the laser that also propagates along $+z$, leading to more energy transfer to the electron. As described in Supplementary Material Eq. (S29), the radiated power depends on the factor $\bm{\beta}=\bm{v}/\bm{c}$. Therefore, increasing the velocity will, in turn, increase the XUV emission. Second, the force generated by the Gaussian pulse remains nearly monochromatic, i.e., the same center frequency, whereas the toroidal pulse induces forces with varying center frequencies, indicating a stronger deviation from simple sinusoidal motion. The large variation in $v_z$ together with the non-monochromatic acceleration leads to enhanced radiation at higher emission frequencies.

\begin{figure}[h]
    \centering
\includegraphics[width=1\linewidth]{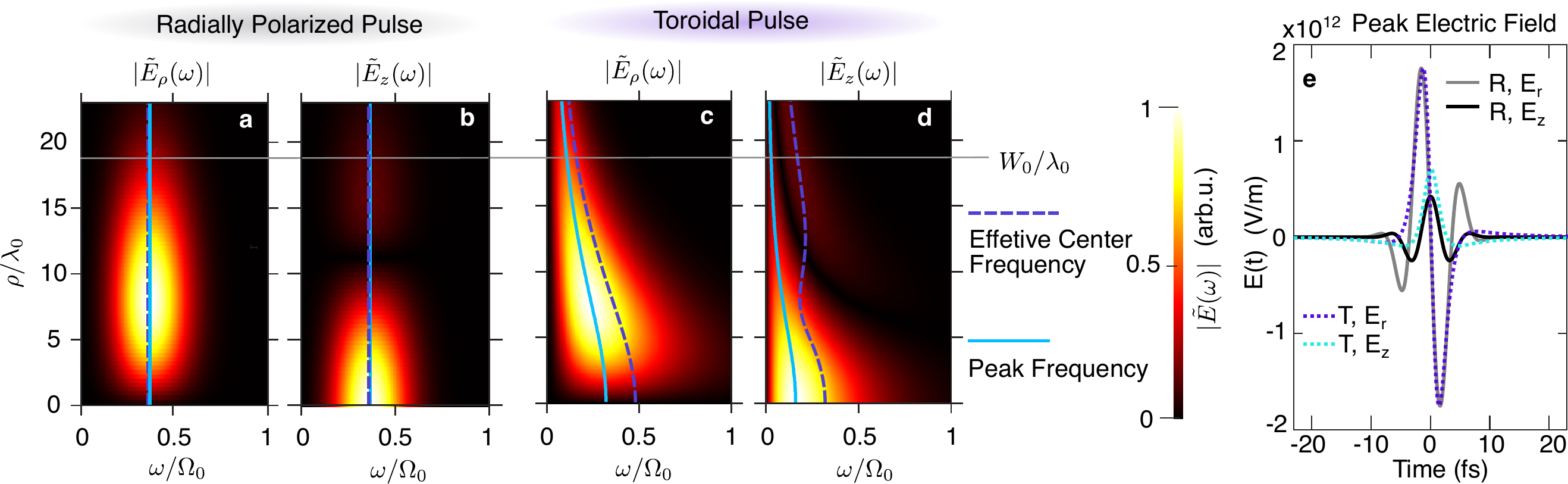}
\caption{Spatio-spectral field distribution for both the $\rho$ (panel \textbf{a,c}) and $z$ (panel \textbf{b,d}) components. The blue curves denote the peak frequency (spectral maximum). The purple dashed curves mark the effective center frequency obtained from Eqs.~(\ref{eq:w_r}). The horizontal line at $W_0/q_1$ indicates the beam size. We have chosen $q_1=\lambda_0=0.8$\textmu m and $\Omega_0=2\pi c/\lambda_0$. The effective center frequency $\omega_c\approx0.31\Omega_0$. The peak field for all the components is presented in panel \textbf{e}.}\label{fig3}
\end{figure}

The root of this asymmetry arises from the non-separable spatiao-temporal structure of the toroidal pulse as mentioned earlier. The electric fields of the toroidal pulse have a markedly asymmetric spectral distribution, accompanied by a pronounced high-frequency tail as shown in Fig.~\ref{fig3}, where the spatio-spectral distributions of the $\rho$ and $z$ components are presented. Fig.~\ref{fig3} \textbf{a,b} correspond to the radially polarized pulse, whereas Fig.~\ref{fig3}\textbf{c,d} depict the toroidal pulse. The Gaussian pulse coincides with that shown in Fig.~\ref{fig3}\textbf{a}. During the interaction, the electron travels transversely to different positions within the toroidal pulse and experiences forces with varying frequencies, leading to increased nonlinearity.

The peak electric fields of all the field components are presented in Fig.~\ref{fig3}\textbf{e}. It can be seen that the $E_r$ component of the radially polarized and toroidal pulses is almost identical around the major peaks. Our analysis shows that the $E_z$ component of the toroidal pulse primarily affects the emission intensity, while only moderately influencing its spectral extent (Supplementary Material Fig.S5). 
Specifically, we define $\tilde{E}_\rho(\omega,\rho)=\text{FT}[E_\rho(t,\rho)]$ and $\tilde{E}_z(\omega,\rho)=\text{FT}[E_z(t,\rho)]$, where FT denotes the Fourier transform. The purple dashed curves indicate the effective center frequency, obtained from Eqs.~(\ref{eq:w_r}). The peak frequency, defined as the maximum of the spectrum, is indicated by the blue curves. For the radially polarized pulse, the peak and effective center frequencies are nearly identical for the chosen parameters. In contrast, for the toroidal pulse, the effective center frequency is consistently higher than the peak frequency, reflecting the spectrum's asymmetric long tail. The horizontal line at $W_0/\lambda_0$ indicates the beam size.

\begin{figure}[h]
    \centering
\includegraphics[width=0.3\linewidth]{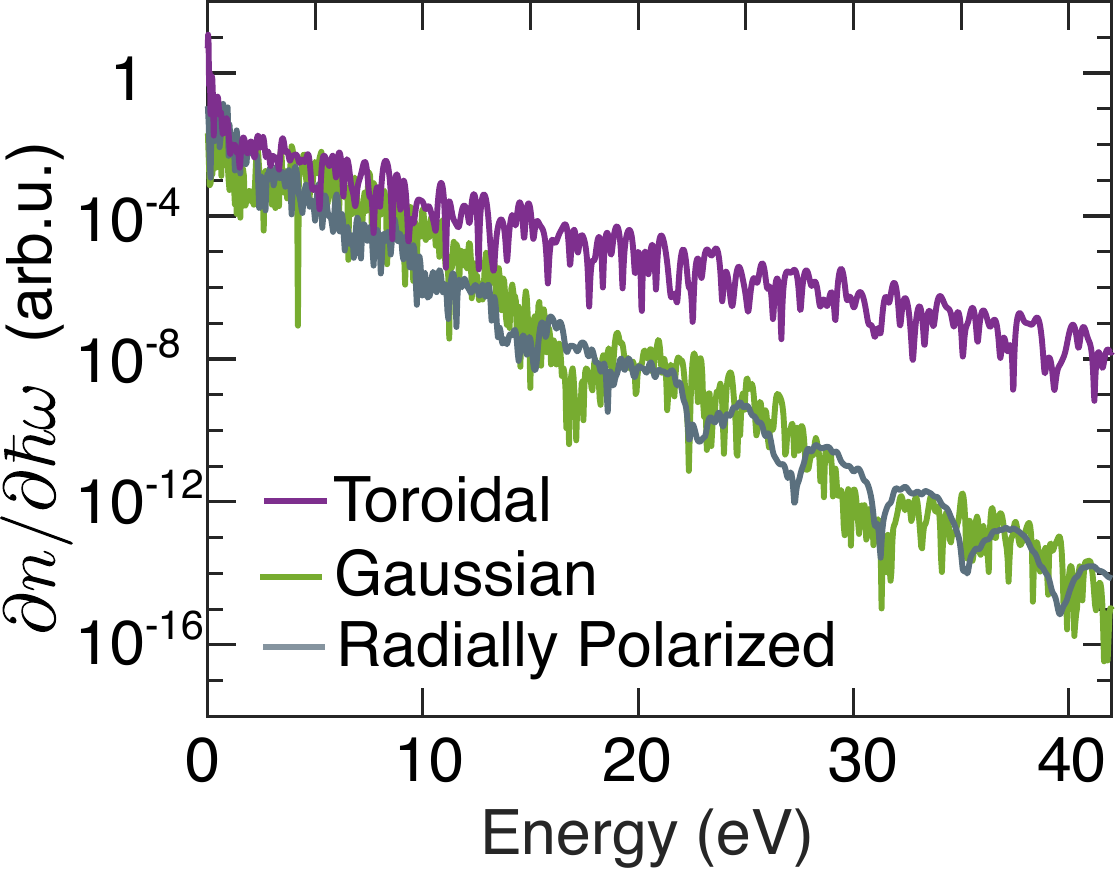}
\caption{The differential photon number per energy  $\partial n/\partial \hbar \omega$ from 500 randomly distributed electrons. }\label{fig4}
\end{figure}

All analyses above are performed for a single electron. In realistic scenarios, acceleration involves many electrons, and the cumulative emission from the ensemble ultimately governs the observed response. To fulfill this aspect, we compute the emission from 500 non-interacting electrons, each with zero initial velocity and initial positions randomly distributed within the beam cross-section, i.e., $x_0, \,y_0 \sim [-W_0, W_0]$ and $z_0\sim [-2c\tau_c,2c\tau_c]$.

Figures~\ref{fig4} shows the differential photon number per energy  $\partial n/\partial \hbar \omega$ over the range $0$–$40$,eV. We can see that the toroidal pulse yields emission in the XUV range that is 2–4 orders of magnitude stronger compared to the Gaussian and radially polarized pulses.

\section{Discussion and Conclusion}
Our study demonstrates that toroidal pulses provide a fundamentally different mechanism for XUV generation compared with conventional laser pulses. Owing to their intrinsically non-separable spatiotemporal field structure, toroidal pulses drive highly asymmetric electron acceleration dynamics that cannot be achieved using standard Gaussian or radially polarized beams. This unique acceleration process leads to substantially enhanced net velocity gain and yields XUV emission that is 2–4 orders of magnitude stronger than that generated by conventional pulses under comparable conditions.

The strong enhancement observed here highlights the importance of spatiotemporal field engineering in extreme nonlinear light–matter interactions.  Beyond XUV generation, toroidal pulses may open new opportunities for compact attosecond light sources, particle acceleration, and nonlinear radiation control. Since toroidal pulses naturally provide tightly confined single-cycle electromagnetic fields, they offer additional advantages for ultrafast electron manipulation. Overall, this work establishes toroidal pulses as a promising tool for advanced strong-field and ultrafast light–matter interaction studies.
\,\\
\section{Methods}
\subsection{Laser Pulse Parameters}
Note that, throughout this work, the pulse energy, beam size, effective central frequency $\omega_c$, and effective pulse duration $\tau_c$ are held constant across all three pulse types—Gaussian, toroidal, and radially polarized (see Supplementary Material Fig.S3).
The energy, valid for single-cycle pulses, of the three laser pulses discussed can be written as (Supplementary Material Section IV.) \cite{hellwarth1996focused,ziolkowski1989localized}\begin{align}
&U_\text{Toroidal}=\frac{\pi^2\varepsilon_0f_0^2(q_1+q_2)}{2q_1^3q_2^3},\\
&U_\text{Gaussian}=c\varepsilon_0\tau_c E_{0,G}^2 W_{0,G}^2\left(\sqrt{\frac{\pi}{2}}\right)^{\,3}\left\{1+\frac{1}{\omega_c^2\tau_c^2}\left[1-\sgn\left[\cos(2\phi_\text{CEP})\right]\exp\left(\frac{-\omega_c^2\tau_c^2}{2}\right)\right]\right\}\\
&U_\text{Radial}=c\varepsilon_0\tau_c E_{0,R}^2W_{0,R}^2 \left(
    \sqrt{\frac{\pi}{2}}\right)^3\left\{1+\frac{1}{\omega_c^2\tau_c^2}\left[1-\sgn\left[\cos(2\phi_\text{CEP})\right]\exp\left(\frac{-\omega_c^2\tau_c^2}{2}\right)\right]\right.\nonumber\\
    &\left.+\frac{2}{W_{0,R}^2(\omega_c/c)^2}\left[1+\frac{1}{\tau^{2} \omega^{2}}\left(1 +\sgn\left[\cos(2\phi_\text{CEP})\right]\exp{\left(\frac{-\tau^2\omega_0^2}{2}\right)}\right)\right]\right\},  
\end{align}
where $\varepsilon_0$ is the vacuum permittivity, $c$ is the speed of light, $\phi_\text{CEP}$ is the carrier envelope phase. $\phi_\text{CEP}=0$ and $\pi/2$ values correspond to cosine- and sine-type temporal pulses, respectively. To maintain a comparable beam size for all three cases, we have chosen 
\begin{equation}
 W_{0,G}=W_0, \, W_{0,R}=W_0\exp{\left(-\frac{1}{2}\right)}, q_2=\frac{W_0^2}{2q_1},
\end{equation}
The parameters $E_{0,G}$, $f_0$, $E_{0,R}$ are related to electric field strength and  are adjusted accordingly to make sure all the pulses have the same energy, i.e.  
\begin{equation}
    U_\text{Toroidal}= U_\text{Radial}=U_\text{Gaussian}.
\end{equation}
With the above conditions, all three pulses have comparable peak field strength $E_0\sim 2\times 10^{12}\,\text{V}/$m (see Supplementary Material Fig.S3), corresponding to a dimensionless amplitude $a_0=eE_{0}/(m\omega_c c)\sim 1.6$, where $e=|e|$ is the elementary charge, $m$ is the mass of the electron \cite{esarey1993nonlinear,lee2003relativistic}.

\subsection{Effective Center Frequency and Pulse Duration of the Toroidal Pulse}
For the toroidal pulse, we have used $q_1=\lambda_0=0.8$\textmu m, $q_2={W_0^2}/{(2q_1)}$, where $q_1\ll q_2$, and $W_0=15$\textmu m is the beam size. The effective central angular frequency at different $\rho$ is defined as 
\begin{equation}\label{eq:w_r}
    \omega_{i}(\rho)=\frac{ \int_0^\infty  \omega|\tilde{E}_i(\omega,\rho)|d\omega }{ \int_0^\infty  |\tilde{E}_i(\omega,\rho)|d\omega},
\end{equation}
where $i\in \{z, \rho\}$ represents the vector component, $\tilde{E}_i(\omega,\rho)$ is the spectral i.e., the Fourier transform of the electric field ${E}_i(t,\rho,z=0)$ (analytical expressions can be found in Supplementary Material Section IV. B). The effective angular frequency of the radial component of the toroidal pulse can be written as
\begin{equation}
\omega_\rho(\rho)=\frac{3c(q_1+q_2)}{q_1q_2+\rho^2}.
\end{equation}
Since $\tilde{E}_z(\omega,\rho)$ changes sign as a function of both $\omega$ and $\rho$, the effective angular frequency does not reduce to a simple analytical form. Consequently, we have obtained $\omega_z(\rho)$ using Eq.(\ref{eq:w_r}) numerically. In addition, we have obtained the spectral bandwidth (FWHM) at each position $\rho$ by numerically finding two frequencies $\omega_1$ and $\omega_2$ that satisfy $\tilde{E}_i(\omega_{1,2},\rho)=\frac{1}{2}\max{\left[|\tilde{E}_i(\omega,\rho)|\right]}$. The corresponding bandwidth is then defined as
\begin{equation}
\Delta_{i}(\rho)=|\omega_2(\rho)-\omega_1(\rho)|. \label{eq:BW_omega}
\end{equation}
 By denoting $I_i(\rho)=\int_0^{\infty} |\tilde{E}_i(\omega,\rho)|^2d\omega$, where $i\in \{z, \rho\}$, we obtain the effective center frequency $ \omega_{c}$ and the bandwidth $\Delta_{c}$ of the entire toroidal beam as
\begin{align}
&\omega_{c}=\frac{\int_0^{W_0} \rho\sum_i I_i(\rho)\omega_i(\rho) d\rho}{\int_0^{W_0} \rho\sum_i I_i(\rho) d\rho}\approx0.31\Omega_0,\label{eq:w_c}\\
&\Delta_{c}=\frac{\int_0^{W_0} \rho\sum_i I_i(\rho)\Delta_i(\rho) d\rho}{\int_0^{W_0} \rho\sum_i I_i(\rho) d\rho}\approx0.34 \Omega_0,\label{eq:bw_c}
\end{align}
where $\Omega_0=2\pi c/\lambda_0$. The upper limit $W_0$ is chosen instead of $\infty$ over $\rho$ because $\rho<W_0$ already contains more than $93\%$ of the total energy with our choice of parameters. In addition, at the outer part of the beam, $\tilde{E}_z(\omega,\rho)$ contains multiple oscillations. This introduces discontinuity in Eqs.(\ref{eq:w_c},\ref{eq:bw_c}). As a result, $\rho>W_0$ regime is neglected. Using this effective bandwidth, we define the Gaussian pulse duration $\tau_c$ for an electric field envelope of the form $\exp(-t^2/\tau_c^2)$ to be
\begin{equation}  \tau_c=4\sqrt{\log{(2)}}/\Delta_{c}\approx 4.1\,\text{fs}.
\end{equation}

\section{Acknowledgments}
L.W. would like to show heartfelt gratitude to the HPC support from Digital Research Alliance of Canada; to thank Milton and Rosalind Chang Pivoting Fellowship from the Optica Foundation; the Ministry of Education Singapore, Academic Research Fund Tier 2 (T2EP50125-0015); and X-lites (NSF Award 2411691, U.S.A.). T.B. thanks NSERC. Z.C. thanks Air Force Office of Scientific Research (FA9550-18-1- 0223) and the Canada Excellence Research Chair program.
\section{Data Availability}
The numerical package: Induced Structured Intense Lasers Driven Ultra-fast Radiation (Isildur), is available on the GitHub repository \url{httpxx...}

\bibliographystyle{apsrev4-1} 
\bibliography{my_bib} %
\end{document}